\pdfoutput=1

\documentclass[11pt]{article}

\usepackage[final]{ACL2023}
\usepackage{times}
\usepackage{latexsym}
\usepackage[T1]{fontenc}

\usepackage[utf8]{inputenc}

\usepackage{microtype}

\usepackage{inconsolata}
\usepackage{xcolor}
\usepackage{multirow}
\usepackage{adjustbox}
\usepackage{booktabs} 
\usepackage{enumitem}
\usepackage{tabularray}
\usepackage{pifont}
\usepackage[most]{tcolorbox}
\usepackage{xspace}
\usepackage{longtable,tabu}
\usepackage{algorithm}
\usepackage{algpseudocode}
\usepackage{soul}
\usepackage{bm}

\usepackage{listings}
\usepackage[utf8]{inputenc}

\definecolor{codegreen}{rgb}{0,0.6,0}
\definecolor{codegray}{rgb}{0.5,0.5,0.5}
\definecolor{codepurple}{rgb}{0.58,0,0.82}
\definecolor{backcolour}{rgb}{0.95,0.95,0.92}
 
\lstdefinestyle{mystyle}{
commentstyle=\color{codegreen},
keywordstyle=\color{magenta},
stringstyle=\color{codepurple},
basicstyle=\scriptsize\ttfamily,
breakatwhitespace=false,         
breaklines=true,                 
captionpos=b,                    
keepspaces=true,                 
showspaces=false,                
showstringspaces=false,
showtabs=false,                  
tabsize=2,
}
\newcommand*{\methodnamews}{\textsc{RepoExec}\@\xspace}

\title{On the Impacts of Contexts on Repository-Level Code Generation}

\author{Nam Le Hai$^{1,2}$, Dung Manh Nguyen$^1$, Nghi D. Q. Bui$^1$  \\
         $^1$FPT Software AI Center, Vietnam \\ 
         $^2$Hanoi University of Science and Technology, Viet Nam \\
         \texttt{namlh@soict.hust.edu.com, dungnm31@fpt.com,  bdqnghi@gmail.com}}
         
\newcommand{\cmark}{\ding{51}}%
\newcommand{\xmark}{\ding{55}}%

\makeatletter
\newcommand*{\algrule}[1][\algorithmicindent]{%
  \makebox[#1][l]{%
    \hspace*{.2em}
    \vrule height .75\baselineskip depth .25\baselineskip
  }
}

\newcount\ALG@printindent@tempcnta
\def\ALG@printindent{%
    \ifnum \theALG@nested>0
    \ifx\ALG@text\ALG@x@notext
    \else
    \unskip
    \ALG@printindent@tempcnta=1
    \loop
    \algrule[\csname ALG@ind@\the\ALG@printindent@tempcnta\endcsname]%
    \advance \ALG@printindent@tempcnta 1
    \ifnum \ALG@printindent@tempcnta<\numexpr\theALG@nested+1\relax
    \repeat
    \fi
    \fi
}
\patchcmd{\ALG@doentity}{\noindent\hskip\ALG@tlm}{\ALG@printindent}{}{\errmessage{failed to patch}}
\patchcmd{\ALG@doentity}{\item[]\nointerlineskip}{}{}{} 
\makeatother

\begin{document}
\maketitle

\begin{abstract}

CodeLLMs are widely used for code generation, yet their ability to handle repository-level dependencies remains underexplored. We introduce \textbf{\methodnamews}, a benchmark for evaluating repository-level code generation, focusing on executability, functional correctness, and dependency utilization. Our study evaluates 18 models, revealing that retaining full dependency context yields the best performance, while smaller context sizes can be misleading. Pretrained LLMs excel in correctness but often reimplement dependencies, while instruction-tuned models better utilize dependencies but sometimes introduce unnecessary complexity. We propose an instruction-tuning dataset that improves dependency handling and introduce a new metric, \textit{Dependency Invocation Rate (DIR)}, to measure context utilization. Experiments show that instruction-tuned models improve DIR by over 10\%, and multi-round debugging further enhances both correctness and dependency use. \methodnamews provides a comprehensive framework to advance CodeLLMs for real-world applications. The dataset and source code are available at~\url{https://github.com/FSoft-AI4Code/RepoExec}.

\end{abstract}

\section{Introduction}
\begin{figure*}[t!]
  \centering
  \small
  \includegraphics[width=0.95\textwidth]{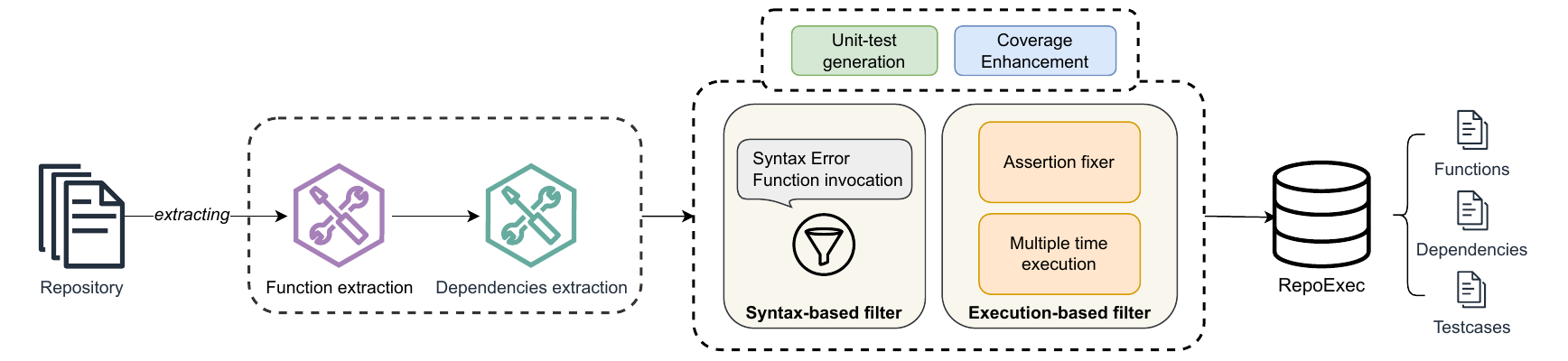}
  \caption{Data Processing Pipeline of \methodnamews}
  \label{fig:data_pipeline}
\end{figure*}

Code Large Language Models (CodeLLMs) have emerged as powerful tools for assisting with coding tasks~\cite{wang2021codet5,wang2023codet5+, feng2020codebert, allal2023santacoder, li2023starcoder, lozhkov2024starcoder, guo2024deepseek, pinnaparaju2024stable, zheng2024opencodeinterpreter, roziere2023code, nijkamp2022codegen, luo2023wizardcoder, xu2022systematic, bui2023codetf, manh2024codemmlu, liu2024deepseek}. While these models excel at generating code from natural language requirements or completing individual lines of code, their application in real-world, professional software development scenarios presents more complex challenges. A critical aspect of this complexity lies in leveraging relevant \textbf{\textit{contexts/dependencies}} (e.g., function calls, imports, class hierarchy) across an entire software repository, which raises two pivotal questions. First, to what extent are the retrieved contexts accurate and relevant, rather than potential noise in the input? Second, how effectively do LLMs process and incorporate the provided dependencies into their generated code? These inquiries are central to understanding the capabilities and limitations of CodeLLMs in repository-level code generation, where completing a single line of code might require making API calls to functions within the same file (in-file context) or across different files (cross-file context).

Existing repository-level code generation benchmarks, such as RepoBench~\cite{liu2023repobench}, RepoCoder~\cite{zhang2023repocoder}, CrossCodeEval~\cite{DBLP:conf/nips/DingWADTJRNBRX23}, CoCoMIC~\cite{ding-etal-2024-cocomic-code}, and DevEval~\cite{li2024deveval}, have advanced the evaluation of code generation at a repository level. However, they exhibit key limitations:
(1) \textit{Lack of an executable environment}, leading to dependence on match-based metrics that fail to assess functional correctness adequately; (2) \textit{Insufficient control over unit test quality}, reducing the data pipeline scalability and robustness of the evaluation; and (3) \textit{Overemphasis on functional correctness metrics like pass@k}, which is inadequate for comprehensive repository-level code generation evaluation. In real-world development, code often needs to call predefined modules (functions, classes, or variables) in the repository to align with developer intent. While LLMs may generate code that passes tests, it can result in inefficient implementations or unnecessary duplication of predefined functions, leading to issues like technical debt and code smells \citep{maldonado2015detecting, sierra2019survey, santos2018systematic}.

To address these gaps, we propose \methodnamews, a novel benchmark designed to ensure both functional correctness and effective dependency utilization. It offers an executable environment for reliable functional evaluation and incorporates a mechanism to generate high-coverage test cases, enhancing the robustness of functional correctness assessment. Additionally, we also propose \textit{Dependency Invocation Rate (DIR)}, a metric that evaluates the extent to which generated code leverages predefined dependencies, offering a more comprehensive measure of model performance beyond merely passing test cases.

Our experiments with \methodnamews offer several key insights into repository-level code generation. Firstly, 
given varying context levels (full, medium, small), which correspond to the information provided for dependencies—such as full implementation, docstrings, and function signatures—we observe that models achieve the best performance when given full context dependencies while also showing potential with smaller contexts. Secondly, pretrained LLMs and instruction-tuned LLMs exhibit different strengths: pretrained LLMs generally excel in pass@k metrics, while instruction-tuned LLMs perform better on the Dependency Invocation Rate (DIR). This suggests that while pretrained LLMs can generate runnable code, they often struggle with properly utilizing dependencies, which means the code may pass tests but might not be correctly implemented in a repository-level context. On the other hand, instruction-tuned LLMs make better use of contextual information. These findings highlight the importance of using both pass@1 and DIR for a more holistic evaluation, emphasizing the need to improve both functional correctness and context utilization in CodeLLMs.

Furthermore, we investigate two methods to enhance repository-level code generation: multi-round debugging and instruction tuning with dependency contexts. Multi-round debugging with test execution, particularly with models like GPT-3.5 and WizardCoder, significantly boosts both pass@1 and DIR after three rounds. Additionally, fine-tuning on our dependency-enhanced dataset improves these metrics while reducing computational costs. These findings highlight the benefits of executable code testing, instruction tuning, and iterative debugging for improving CodeLLM performance and managing code dependencies more effectively. Besides, it also demonstrates that better context utilization leads to higher pass@k scores (detailed analysis provided in Sections \ref{sec:debug}, \ref{sec:inst_tune} and Appendix \ref{appendix_debug}). In summary, our contributions are as follows:

\begin{enumerate}[leftmargin=*]
    \item We introduce a novel \textit{evaluation paradigm} for repository-level code generation, assessing both functional correctness and quality factors such as maintainability and adherence to clean code principles through efficient dependency use.
    
    \item We present the Dependency Invocation Rate (DIR), a novel metric that measures the proportion of provided dependencies successfully incorporated into the generated code. This metric helps gauge the models' understanding and utilization of dependencies and shows a strong correlation with functional correctness.

    \item We introduce \methodnamews, a novel benchmark that aligns with our evaluation paradigm, addressing the gaps in existing benchmarks and offering a comprehensive assessment of code generation quality. Additionally, an effective pipeline is introduced within \methodnamews from dependency extraction, dynamically generate high-coverage test cases and automatic evaluation with execution and code dependencies. Our pipeline offers practical usage and scalability for the community.

    \item We release a tool named \texttt{pydepcall} to extract dependencies for all functions within any repository, providing a practical use for advancing research in this domain.

    \item Our experiments reveal key insights into CodeLLMs' performance in repository-level code generation. While foundation models show high initial accuracy, instruction-tuned models excel in dependency management. Additionally, multi-round debugging tests further improve performance, enhancing dependency management. Notably, a strong correlation between pass@1 and DIR underscores that better context utilization leads to more functionally correct code.
    
\end{enumerate}

\section{Related works}
\label{sec:related_work}
Coding-related tasks have been crucial for assessing the performance of Large Language Models (LLMs), with code generation emerging as a primary focus \citep{chen2021evaluating, li2023starcoder, jiang2024mixtral, touvron2023llama, roziere2023code, xu2022systematic, allal2023santacoder, nijkamp2022codegen, phan2024repohyper, to2023better}. Early benchmarks have been introduced to address this issue \cite{yin2018learning, iyer2018mapping, manh-etal-2023-vault, chen2021evaluating, austin2021program, hendrycksapps2021}; however, they often had limited scope or employed weak evaluation approaches. Some benchmarks \cite{yin2018learning, iyer2018mapping, manh-etal-2023-vault} exhibit domain diversity akin to real-world applications; however, their evaluation methodologies are constrained to match-based metrics, thereby decreasing the reliability of these benchmarks \cite{chen2021evaluating}. Meanwhile, benchmarks with reliable evaluation approaches such as HumanEval \cite{chen2021evaluating}, MBPP \cite{austin2021program} and APPS \cite{hendrycksapps2021} often entail limitations to specific domains like competitive programming. Recently, there have been efforts to extend the domains of generation tasks in various benchmarks. For instance, ExeDS \cite{huang2022execution} focuses on data science code generation, while ODEX \cite{wang2022execution} serves as an open-domain benchmark for code generation. Besides, all the benchmarks mentioned primarily focus on standalone function generation, lacking consideration for cross-contextual and dependency invocation scenarios.

Several recent studies have introduced frameworks and benchmarks for repository-level code generation \citep{ding-etal-2024-cocomic-code, DBLP:conf/icml/ShrivastavaLT23, DBLP:conf/nips/DingWADTJRNBRX23, liao2023context, liu2023repobench}. These studies closely align with real-world scenarios, underscoring the importance of cross-contextual considerations. However, these works are still limited to match-based evaluation methods. A recent study \cite{li2024deveval} introduced the DevEval benchmark for code generation within repository contexts, evaluating performance based on extracted tests available in the repository. These benchmarks primarily focus on assessing the functional correctness of generated outputs but have not extensively investigated the correctness in conjunction with the ability to utilize dependency contexts.

\section{Evaluation Paradigm}

In this section, we outline our paradigm to achieve a more robust and comprehensive evaluation of repository-level code generation. Our paradigm encompasses two key attributes: Functional correctness and Dependency utilization.

\paragraph{Functional correctness:} This evaluation criterion ensures that the code accurately performs its intended tasks and requirements. Specifically, it typically involves using test cases to compare the execution output of the generated code against the expected output for a given input. This criterion has been widely employed for evaluating code generation in numerous studies \citep{chen2021evaluating, austin2021program, zhuo2024bigcodebench, li2024deveval}. We follow the well-known automatic metric \textit{Pass@k} \cite{chen2021evaluating} to measure the functional correctness of generation outputs. 

\paragraph{Dependency utilization:} Functional correctness alone cannot fully capture code quality, as it may overlook poor implementations or redundancy. Match-based metrics like BLEU and edit similarity evaluate alignment between generated and reference code at the token level. However, not all tokens impact code quality equally, and alternative implementations can maintain quality despite low similarity scores. Tokens representing called dependencies—such as packages, functions, variables, and classes within the repository—imply effective use of human intent for efficient implementation. Ignoring these dependencies may suggest workaround implementations misaligned with human intent, leading to increased verification and maintenance costs.
To assess the models' ability to utilize provided dependencies in accordance with human intent, we introduce the \textit{Dependency Invocation Rate (DIR)}. This metric represents the percentage of invoked dependencies out of the total number of dependencies provided. Let $D_g$ denote the set of identifiers in the generated output, and $D_s$ denote the set of provided dependencies extracted from the solution. The Dependency Invocation Rate (DIR) is calculated as follows:
\begin{equation}
\text{DIR} = \dfrac{|D_g \cap D_s|}{|D_s|} \nonumber
\end{equation}
\textit{A higher DIR indicates that the model successfully incorporates a larger proportion of the provided dependencies into the generated code, demonstrating a better understanding of the dependencies' relevance and their intended usage. Conversely, a lower DIR suggests that the model may struggle to identify and utilize the appropriate dependencies, potentially generating code that is less aligned with the human-intent solution.}
In summary, achieving a high-quality solution requires the generated code to excel in both functional correctness and dependency utilization. Otherwise, it indicates a lack of ability to generate correct solutions or suggests poor implementation practices, including technical debt or code smell issues.

\section{Data Collection Pipeline}


\begin{figure*}[t!]
  \makebox[\textwidth][c]{\includegraphics[width=\textwidth]{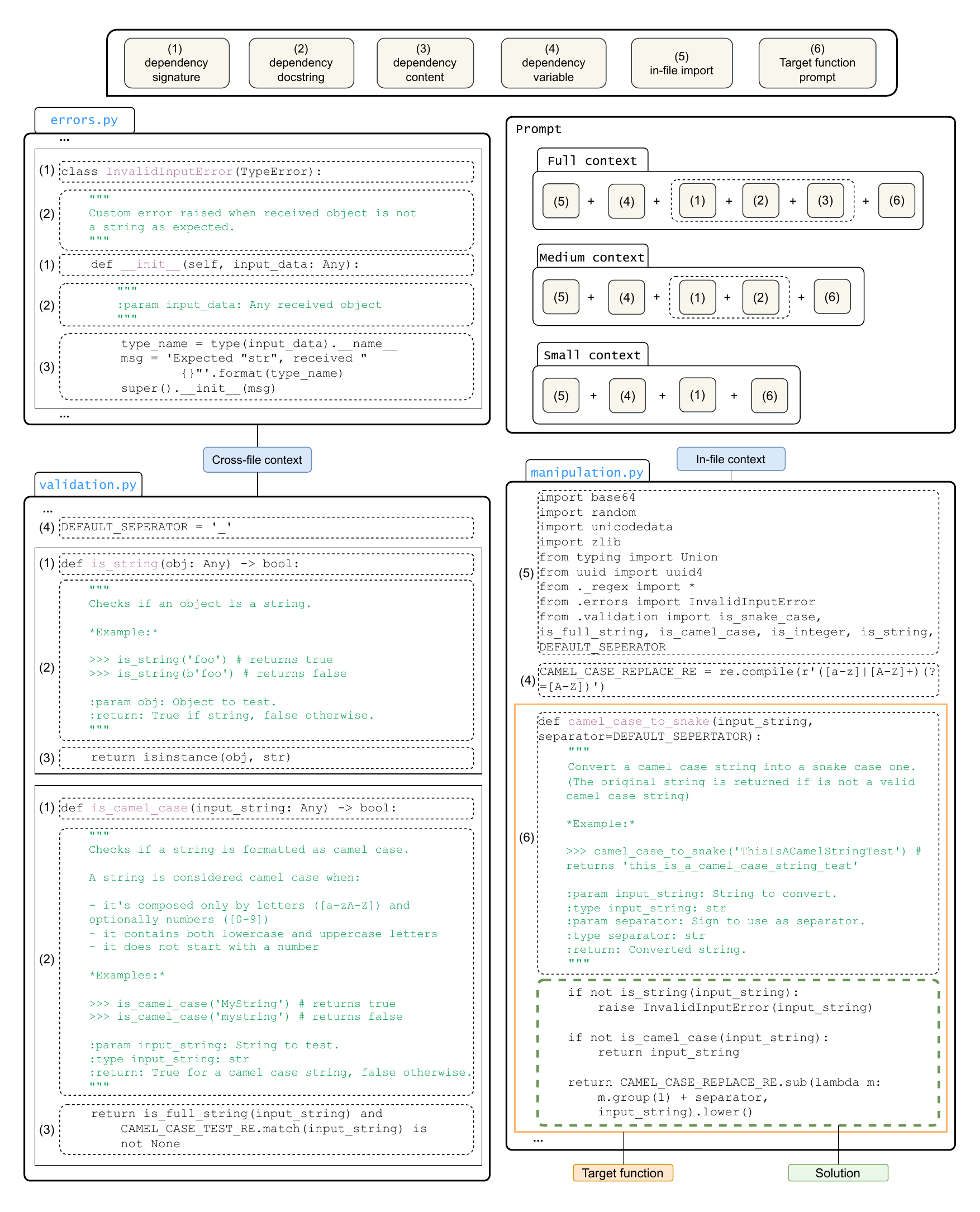}}
  \caption{Illustration of a data instance in RepoExec. The target function signatures and their associated docstrings, which describe the functionality of the functions, are shown in (6). The infile-imports and variable declarations are represented by (5) and (4), respectively. The remaining components, (1), (2), and (3), represent the function and class contexts. Specifically, (1) denotes the class or function signature, (2) may contain the description of the class, and (3) represents the function body of the cross-file function.}
  \label{fig:data_context}
\end{figure*}

Developing an executable benchmark within repository contexts is challenging due to complex setup requirements and frequent lack of clear installation guidelines in repositories. Previous studies \citep{ding-etal-2024-cocomic-code, DBLP:conf/icml/ShrivastavaLT23, DBLP:conf/nips/DingWADTJRNBRX23, liao2023context} have often relied on match-based metrics for evaluation, which may not fully capture code functionality. In addition, test cases are essential for assessing code functionality. However, extracting test cases from repositories \citep{DBLP:conf/emnlp/ZhangCZKLZMLC23, li2024deveval, zhang2024codeagent} often relies on available functions and heuristic rules, limiting adaptability and excluding data without existing tests. For instance, \citealt{li2024deveval} found that over 99\% of functions were discarded due to the absence of suitable tests. We propose a dataset collection pipeline to ensure that repositories can build executable environments and that test cases are automatically generated. 
The complete data pipeline is illustrated in Figure \ref{fig:data_pipeline}, with details of the data sources provided in Appendix \ref{appedix_datasource}.


\begin{table}[ht]
    \footnotesize
    \centering
    \renewcommand{\arraystretch}{1.1}
    \begin{adjustbox}{width=0.49\textwidth}
    \begin{tabular}{l|cc|cc|cc}
    \hline
    & \multicolumn{2}{c|}{\#No problem} & \multicolumn{2}{c|}{\#No testcase} & \multicolumn{2}{c}{Avg tokens} \\
    
     & Cross-file & Total & Avg LC (\%) & Avg & Prompt & Solution\\
    \hline 
    Full & 22.8\% & 355 & 96.25 & 99.45 & 362.92 & 78.46 \\
    Medium & - & - & - & - & 253.05 & - \\
    Small & - & - & - & - & 179.66 & - \\
    \hline
    \end{tabular}
\end{adjustbox}
\caption{Dataset attributes with different levels of contexts. \textbf{Cross-file} refers to the percentage of problems that involve cross-file dependencies. Tree-sitter is used for tokenization. \textbf{AVG LC}: Average Line Coverage.}
\label{tab:dataset_attr}
\end{table}

\subsection{Functions and Dependencies Extraction}
\label{sec:parser_tool}
\paragraph{Function extraction:} We extract functions and their dependencies from repositories, considering only those suitable for function-level code generation, akin to benchmarks like HumanEval \cite{chen2021evaluating} and MBPP \cite{austin2021program}. Using tree-sitter, we parse files into Abstract Syntax Trees (AST) to extract functions, focusing on those with comprehensive docstrings and excluding functions used as entry points or that do not produce verifiable outputs.

\paragraph{Dependencies extraction:} We employ static analysis on call graph to identify dependencies, excluding identifiers that are function parameters or typing objects. Each dependency name is then mapped to its definition within the repository using a repository graph and static analysis tools. We release our tool \texttt{pydepcall} for community usage and provide a brief description in Appendix \ref{sec:appendix_tool}.

For example in Figure \ref{fig:data_context}, for the function \texttt{camel\_case\_to\_snake} in \texttt{manipulation.py}, our process identifies \texttt{CAMEL\_CASE\_REPLACE\_RE} as an in-file dependency and analyzes import statements to track cross-file dependencies from \texttt{errors.py} and \texttt{validation.py}. Dependencies are then parsed and incorporated into the input prompt to ensure comprehensive context for code generation.

\subsection{Test case generation}
\label{sec:testgen_tool}

To overcome the limitation of requiring available tests in the repository for evaluation as in previous works, we leverage large language models (LLMs) to generate test cases automatically. Our proposed approach ensures the correctness of created test cases while also controlling and enhancing test coverage. To execute and validate the generated test cases, it is necessary to configure each repository to create an executable environment. Follow \citealt{lemieux2023codamosa}, we use \texttt{pipreqs}\footnote{\url{https://github.com/bndr/pipreqs}} to identify each project’s dependency packages.

In our test generation process, we conduct two phases corresponding to \textbf{Correctness Control} and \textbf{Coverage Enhancement}. After the test generation process, we exclude samples with a line coverage lower than 40\%, as they are insufficient to accurately assess the correctness of the generated code.

\subsubsection{Correctness Control}
\label{sec:correcness_control}

In this stage, we present the procedure for generating initial test cases and ensure the tests' correctness.
Specifically, we use CodeLlama-13b \cite{roziere2023code} and provide the model with the prompt detailed in Appendix \ref{appendix_testgen}. The first 20 assertions form the test cases, which are then filtered via syntax and execution checks to ensure correctness.

\paragraph{Syntax-based filter:} We filter out tests that present syntax errors during parsing into AST. Additionally, tests that do not invoke the function under test are excluded. For instance, while the statement \texttt{``assert 1''} may pass during execution, it is meaningless and negatively impacts the evaluation.

\paragraph{Execution-based filter:}  We use \texttt{pytest}\footnote{\url{https://github.com/pytest-dev/pytest}} to run the generated tests, discarding those with errors except for \texttt{AssertionError}. If an assertion error occurs, the Assertion Fixer resolves it by executing the test with the target function, extracting the output, and updating the assertion. To handle complex return types (e.g., custom objects), we employ \texttt{pickle}\footnote{\url{https://docs.python.org/3/library/pickle.html}} to store results. Additionally, to address flaky tests that are inconsistent due to randomness, each test is executed 10 times for comparison.



\subsubsection{Coverage Enhancement}
Weak unit tests may allow incorrect implementations to pass \cite{liu2023is}. To address this, we propose a strategy for enhancing test coverage using LLMs. Given the complexity of this task, requiring a strong understanding of code, we utilize GPT-3.5 to improve the quality of test cases. We provide GPT-3.5 with three prompts (see Appendix \ref{appendix_testgen}) to handle challenging scenarios, including edge and corner cases. The initial generated tests serve as few-shot examples. We then extract the newly generated tests and ensure their correctness using our methodology from Section \ref{sec:correcness_control}.
Table \ref{tab:data_comparison} shows a line coverage improvement of about 4\% after enhancement, reaching 96.25\%. The performance gap has also increased to over 5\% (Appendix \ref{appendix_coverage}), indicating greater robustness.

\section{Data Characteristics}
\begin{table}[t]
    \centering
    \renewcommand{\arraystretch}{1.1}
    \begin{adjustbox}{width=0.49\textwidth}
    \begin{tabular}{l|c|c|c}
    \hline
    \multirow{2}{*}{Pipeline} & RepoContext & Test & Test  \\
    & Mining & Generation & Enhancement \\
    \midrule
    RepoCoder & \cmark & \xmark & \xmark  \\
    R2E & \cmark & \cmark & \xmark \\
    CodeBenchGen & \xmark & \cmark & \cmark  \\
    \midrule
    \methodnamews & \cmark & \cmark & \cmark \\
    \hline
    \end{tabular}
\end{adjustbox}
\caption{Comparison of data construction pipelines across different repo-level code generation benchmarks.}
\label{tab:pipeline}
\end{table}

\subsection{Data Formatting}
Figure \ref{fig:data_context} illustrates the input data format used in \methodnamews. We retain import information and append dependencies in the order presented in the import statements, placing the target function signature and description at the end of the prompt. To evaluate the reasoning capability of CodeLLMs in repo-level code generation, we propose three prompt types with varying context lengths:

\begin{itemize}[leftmargin=*]
    \item \textbf{Full-size context}: All contexts, including cross-file and in-file contexts, are preserved to assess the model's ability to navigate and utilize the complete information available in the repository.
    \item \textbf{Medium-size context}: Class and function bodies are removed, while their signatures and docstrings are retained. This tests the model's ability to infer the functionality and usage of dependencies based on their interfaces and documentation, reducing the input context length.
    \item \textbf{Small-size context}: Only the signatures of the dependencies are retained. This tests if CodeLLMs can infer the usage of dependencies in the target function given only the function signatures without docstrings, representing the most challenging case with minimal information.
\end{itemize}

Evaluating the model's performance across these context sizes provides insights into the trade-offs between input context length and the model's reasoning capabilities, helping to understand the optimal balance between providing sufficient information and minimizing input size for effective code generation at the repository level.

We follow \citealt{muennighoff2023octopack} to define 2 types of prompt formats in our evaluation of \methodnamews across LLMs: (1) BasePrompt, which concatenates all contexts with the target functions (Figure \ref{fig:data_context}), and (2) InstructPrompt, which includes specific instructions for the LLMs to follow, utilizing two variations as input formats (further details and examples in Appendix \ref{appendix_dataformat}).
\subsection{Dataset Stastistic}

\begin{table}[t]
    \centering
    \scriptsize
    \renewcommand{\arraystretch}{1.1}
    \begin{adjustbox}{width=0.49\textwidth}
    \begin{tabular}{l|c|c|c|c}
    \hline
    Dataset & \#Samples & RC & TC & LC (\%) \\ \hline
    CoNaLA \cite{yin2018learning} & 500 & \xmark & \xmark & - \\
    CONCODE \cite{iyer2018mapping} & 2,000 & \xmark & \xmark & - \\
    HumanEval \cite{chen2021evaluating} & 164 & \xmark & \cmark, \textbf{H} &  99.43 \\
    MBPP \cite{austin2021program} & 974 & \xmark & \cmark, \textbf{H} &  98.48 \\
    RepoCoder \cite{DBLP:conf/emnlp/ZhangCZKLZMLC23} & 373 & \cmark & \cmark, \textbf{P} & - \\
    CrossCodeEval \cite{DBLP:conf/nips/DingWADTJRNBRX23} & 2,665 & \cmark &  \xmark & - \\
    CoCoMIC \cite{ding-etal-2024-cocomic-code} & 6,888 & \cmark & \xmark & - \\
    DevEval \cite{li2024deveval} & 1,874 & \cmark & \cmark, \textbf{P} & - \\
    \hline
    \methodnamews & 355 & \cmark & \cmark, \textbf{A} & 92.46 \\
    \quad + coverage-enhancement &  & &  & 96.25 \\
    \hline
    \end{tabular}
\end{adjustbox}
\caption{The comparison between popular code generation benchmarks and \methodnamews. For test case, we denote \textbf{H}: Human annotated, \textbf{P}: Pre-existing, \textbf{A}: Automated. \textbf{RC}: Repo-context utilization. \textbf{TC}: Test Cases. \textbf{LC}: Line Coverage}
\label{tab:data_comparison}
\end{table}

\begin{table*}[t]
\centering
\renewcommand{\arraystretch}{1.1}
\begin{adjustbox}{width=\textwidth}
\begin{tabular}{c|l|ccc|ccc|ccc}
\toprule
& \multirow{2}{*}{Model} & \multicolumn{3}{c|}{Full context} & \multicolumn{3}{c|}{Medium context} & \multicolumn{3}{c}{Small context}\\
\cline{3-11}
 & & pass@1 & pass@5 & DIR & pass@1 & pass@5 & DIR & pass@1 & pass@5 & DIR \\ 
\bottomrule
\multicolumn{11}{c}{\textbf{BasePrompt}} \\
\midrule
\multirow{6}{*}{\rotatebox[origin=c]{90}{Pre-models}} 
& CodeLlama-13b-Python \cite{roziere2023code} & 38.65 & \underline{43.24} & 62.26 & \underline{32.96} & \underline{38.33} & 56.38 &  \underline{35.66} & \underline{42.41} & 62.67 \\
& StarCoder \cite{li2023starcoder} & 28.08 & 33.95 & 58.75 & 22.54 & 31.83 & 50.74 & 25.54 & 31.45 & 56.67 \\
& StarCoder2-15b \cite{lozhkov2024starcoder} & 27.77 & 32.60 & 60.57 & 18.70 & 23.28 & 39.28 & 23.27 & 29.78 & 53.49 \\
& Mixtral-8x7B-v0.1 \cite{jiang2024mixtral} & 22.82 & 29.14 & 55.90 & 19.38 & 25.25 & 47.22 & 20.54 & 26.30 & 53.40 \\
& Phi-2 \cite{javaheripi2023phi} & 19.04 & 24.56 & 48.22 & 14.54 & 20.34 & 40.85 & 14.82 & 20.69 & 44.54 \\
& Phi-1 \cite{gunasekar2023textbooks} & 14.99 & 18.38 & 43.17 & 12.48 & 15.42 & 37.45 & 12.54 & 15.96 & 38.75 \\
\cmidrule{2-11}
\multirow{12}{*}{\rotatebox[origin=c]{90}{Inst-models}} 
& DeepSeek-R1 \cite{guo2025deepseek} & \textbf{42.57} & - & 70.86  & - & - & - & - & - & - \\
& DeepSeek-V3 \cite{liu2024deepseek} & \underline{42.00} & - & \underline{80.35}  & - & - & - & - & - & - \\
& Llama 3.1-405B-Instruct \cite{dubey2024llama} &  34.86 & - & 78.81 & - & - & - & - & - & - \\
& GPT-4o & 37.14 & - & \textbf{81.43} & - & - & - & - & - & - \\
& GPT-4o-mini & 30.29 & - & 74.75 & - & - & - & - & - & - \\
& CodeLlama-34b-Python \cite{roziere2023code} & 40.93 & \textbf{49.54} & 68.85 & \textbf{35.92} & \textbf{42.95} & 58.15 & \textbf{39.80} & \textbf{45.79} & 64.23 \\
& WizardCoder-Python-13B-V1.0 \cite{luo2023wizardcoder} & 34.31 & 40.06 & 62.90 & 30.99 & 36.75 & 59.50 & 32.54 & 38.34 & 64.67 \\
& Phind-CodeLlama-34B-v2 & 30.08 & 33.49 & 59.47 & 25.25 & 29.40 & 50.73 & 27.55 & 31.85 & 58.93 \\
& CodeLlama-13b-Instruct \cite{roziere2023code} & 28.56 & 32.67 & 57.09 & 26.25 & 30.72 & 49.48 &  26.73 & 33.50 & 54.53 \\
& GPT-3.5 & 27.27 & 37.69 & 63.59 & 23.15 & 33.94 & 52.79 & 22.59 & 33.63 & 55.22\\
& DeepSeek-Coder-7b-Instruct \cite{guo2024deepseek} & 25.18 & 29.91 & 58.50 & 20.23 & 26.02 & 45.76 & 22.20 & 27.74 & 56.69 \\
& Mixtral-8x7B-Instruct-v0.1 \cite{jiang2024mixtral} & 23.41 & 28.71 & 59.83 & 19.04 & 24.83 & 52.75 & 20.45 & 25.84 & 58.17 \\
\toprule
\multicolumn{11}{c}{\textbf{InstructPrompt}} \\
\midrule
\multirow{9}{*}{\rotatebox[origin=c]{90}{Inst-models}}
& DeepSeek-R1 \cite{guo2025deepseek} & 39.71 & - & 53.78  & - & - & - & - & - & - \\
& DeepSeek-V3 \cite{liu2024deepseek} & 41.71  & - & 62.58  & - & - & - & - & - & - \\
& Llama 3.1-405B-Instruct \cite{dubey2024llama} &  39.43 & - &	73.26 & - & - & - & - & - & - \\
& GPT-4o & 38.00 & - & 68.15 & - & - & - & - & - & - \\
& GPT-4o-mini & 32.29 & - & 53.34 & - & - & - & - & - & - \\
& WizardCoder-Python-13B-V1.0 & 26.20 & 30.68 & 67.32 & 24.56 & 30.25 & \underline{67.54} & 24.70 & 29.56 & \underline{68.34} \\
& CodeLlama-13b-Instruct & 25.66 & 30.82 & 62.04 & 27.44 & 34.11 & 63.33 & 26.73 & 32.43 & 64.85 \\
& GPT-3.5 & 23.82 & 39.10 & 40.55 & 19.62  & 36.03 & 37.48 & 19.00 & 34.00 & 35.28 \\
& Mixtral-8x7B-Instruct-v0.1 & 18.11 & 23.04 & \underline{67.73} & 18.54 & 23.12 & \textbf{69.66} & 15.38 & 20.61 & \textbf{68.86} \\
\bottomrule
\end{tabular}
\end{adjustbox}
\caption{Pass@k and DIR results of various LLMs on \methodnamews. \textbf{Bold scores} indicate the highest, while \underline{Underlined scores} denote the second highest. Pre- and Inst- denote Pretrained and Instruction-tuned, respectively. More results are presented in Table \ref{tab:leaderboard}.}
\label{tab:codegen}
\end{table*}

\paragraph{Comparison to Existing Benchmarks:} Table \ref{tab:data_comparison} compares the details of \methodnamews with existing code generation benchmarks. Benchmarks that exclude execution-based evaluation \cite{yin2018learning, iyer2018mapping, DBLP:conf/nips/DingWADTJRNBRX23, ding-etal-2024-cocomic-code} may gather substantial amounts of data; however, they are inadequate for assessing the quality of the generated code. For HumanEval and MBPP, the majority of problems involve standalone functions, which do not reflect real-world scenarios. Besides, benchmarks that rely on human-annotated and pre-existing test cases \cite{chen2021evaluating, austin2021program, zhang2023repocoder, li2024deveval} are challenging to scale and control the test coverage. Finally, repository-context benchmarks \cite{zhang2023repocoder, DBLP:conf/nips/DingWADTJRNBRX23,ding-etal-2024-cocomic-code, li2024deveval} primarily focus on investigating retrieval modules. For example, RepoCoder analyzes a retriever using a sparse bag-of-words model. Similarly, CrossCodeEval experiments and reports performance using various retrievers such as BM25, UniXCoder, and OpenAI ada. CoCoMic proposed CCFINDER to retrieve cross-file context. DevEval is the closest to our work; however, they compare the performance of the generation model based on different given types of file-level context, which can also align with the different contexts provided by different retrievers. In contrast, our approach emphasizes the generation module's ability to understand and utilize human-provided dependency contexts. 

\paragraph{Comparison to Existing Data Pipeline:} 
Several studies, including RepoCoder \citep{zhang2023repocoder}, R2E \cite{jain2024r2e}, and CodeBenchGen \cite{xie2024codebenchgen}, have introduced data pipelines for repository-level code generation. Table \ref{tab:pipeline} compares our pipeline with these approaches across three key aspects: RepoContext mining, Test Generation, and Test Enhancement. These components are crucial for robust evaluation, as missing any can compromise quality. RepoCoder extracts existing test cases but lacks enhancement, limiting scalability. CodeBenchGen uses LLMs to synthesize function contexts, reducing practicality. Lastly, R2E is the most closely related to our approach, however, it similarly neglects test case enhancement like RepoCoder, which can undermine evaluation robustness (as detailed in Appendix \ref{appendix_coverage}).

\paragraph{Dataset attributes:} Table \ref{tab:dataset_attr} outlines the characteristics of \methodnamews, including the total number of examples, the average number of test cases, and the number of tokens in prompts and solutions.

\section{Experiment}

\subsection{Evaluation Results}
\label{sec:llm_exp}

We evaluated 18 CodeLLMs on \methodnamews and presented the results (pass@1, pass@5, and DIR) in Table \ref{tab:codegen}. We use nucleus sampling with temperature set to 0.2, top-p to 0.95, and 10 outputs generated for all models. For evaluating super-large models requiring paid API access, we utilize greedy decoding and report pass@1 and DIR metrics under the Full Context setting. The results show that retaining the full context of dependencies yields the best performance across all models. Surprisingly, Small-size context proves to be more effective than Medium-size context, which we attribute to the context's input format using BasePrompt, potentially misleading the model into interpreting the dependency functions as few-shot examples. Using the Small-size context results in a fair decrease in performance compared to the Full context while effectively reducing the input length.

DeepSeek-R3 achieves the highest pass@1 rate among the evaluated models. Additionally, BasePrompt and pretrained models exhibit superior effectiveness over instruction-tuned models in terms of functional correctness (pass@k) when comparing models of similar size on \methodnamews. However, our analysis reveals limitations in both types of models. Additional discussions and examples are provided in Appendix \ref{appendix_llms_exp}.

\begin{enumerate}[leftmargin=*]
\item Instruction-tuned LLMs demonstrate a higher capacity for utilizing given dependencies than foundation LMs, sometimes enabling them to address edge or corner cases that foundation models have overlooked.
\item Despite the high DIR, instruction-tuning LMs may not utilize dependencies correctly and frequently produce overly complex code, leading to incorrect solutions.
\item Pretrained LLMs generate functionally correct code but often fail to effectively utilize the provided dependencies, occasionally reimplementing dependencies already present in the given context. This leads to redundancy and potentially creates technical debt or code smell issues \citep{maldonado2015detecting, sierra2019survey, santos2018systematic, hai2024improving}.
\end{enumerate}

These issues can lead to a low-quality codebase, requiring substantial human effort for reviewing and fixing, which may even exceed the effort needed to write the code from scratch. Moreover, the performance of current advanced LLMs highlights that DeepSeek models, particularly DeepSeek-R1 and DeepSeek-V3, demonstrate strong overall capabilities, effectively balancing accuracy and dependency management. Among smaller models, GPT-4o-mini achieves the highest DIR (74.75) but underperforms in pass@1 (30.29), emphasizing it often invokes dependencies incorrectly despite its emphasis on handling them. Models with high pass@k and DIR scores, such as DeepSeek-V3 (80.35) and Llama 3.1 (73.26), excel in accurately invoking dependencies, which is crucial for real-world code generation tasks. Finally, \methodnamews remains challenging to existing models, as evidenced by consistently low performance, underscoring the need for more advanced systems.

\begin{table}[t]
    \scriptsize
    \centering
    \renewcommand{\arraystretch}{1.1}
    \begin{adjustbox}{width=0.49\textwidth}
    \begin{tabular}{c|c|c|c}
    \hline
    Round & GPT-3.5 & WizardCoder & CodeLlama-13b-Python \\
    \hline

    \hline 
    0 & 27.04 & 34.37 & 39.44\\
    1 & 36.34 & 40.85 & 39.44 \\
    2 & 40.00 & 41.69 & 39.44 \\
    3 & 41.97 & 42.54 & 39.44\\
    \hline
    \end{tabular}
\end{adjustbox}
\caption{Pass@1 scores of various models across three rounds of debugging. Round 0 represents the initial generation stage.}
\label{tab:debug}
\end{table}

\subsection{Generation with Multi-round Debugging}
\label{sec:debug}

In this section, we examine the models' self-refinement capabilities in enhancing generation performance. We provide the models with error output logs and ask them to fix the errors. We experiment with WizardCoder, GPT3.5, and CodeLlama-13b-Python. The number of rounds to debug is set to 3 and the input template is presented in Appendix \ref{appendix_debug}. In this experiment, we employ a greedy search algorithm to generate only a single output. 

Table \ref{tab:debug} shows the improvement across three rounds of debugging in various models. We observe that GPT-3.5 and WizardCoder demonstrate a high capacity for debugging with improvement of over 10\% and 7\% in pass@1, respectively, while CodeLlama fails to take advantage of this process. Additionally, the DIR has also shown a significant improvement (over 7\%) after three rounds of debugging in these two instruction models (Figure \ref{fig:instruction_debugrs}). These findings indicate a promising approach using self-refinement with debugging for code generation, which can enhance both the correctness and the utilization of given dependencies.

\subsection{Instruction-tuning with Code Dependencies}
\label{sec:inst_tune}

While the multi-round debugging experiment has demonstrated effectiveness in leveraging given dependencies to provide correct solutions (Section \ref{sec:debug}), it requires a strong model to generate good test cases and can be time-consuming due to the repeated generation and execution of code and test cases.
To address these challenges, we propose an instruction-tuning dataset for fine-tuning base LLMs. We collected the 1,555 most-starred repositories from 2018 onward and extracted functions with their corresponding dependencies, following the procedure outlined in Section \ref{sec:parser_tool}. We obtained 154,818 functions, of which 57,746 samples include docstrings. We use 50K samples with docstrings and applied instruction prompts, while 80K samples adhered to the raw code format (Full context). Recognizing the potential of the Small context format, we allocated the remaining 20K samples to follow this structure. We fine-tuned Phi-2, StarCoder, StarCoder2, and CodeLlama-13b-Python models for 5 epochs with LoRa \cite{hu2021lora} and used 10\% of the training data as the validation set to select the best checkpoint.

Table \ref{tab:dep_inst_tuning} illustrates the efficacy of our training data. All 4 models demonstrate improvements in both Pass@1 and DIR after instruction tuning. Specifically, there is a slight increase of around 1\% in Pass@k for all models, while DIR shows a significant improvement, reaching the highest scores (over 70\%) compared to other models after tuning with our dataset. Notably, performance improves significantly with small context, matching results from full context, enabling more efficient processing and reducing computational costs. In summary, our instruction-tuning method enhances the model's ability to utilize dependencies and ensure functional correctness. While multi-round debugging (Section \ref{sec:debug}) is more effective, instruction-tuned models rely on single-turn generation, making them more practical and efficient. \textit{To facilitate open research in this domain, we will publicly release this dataset}.

\begin{table}[t]
\scriptsize
\centering
\renewcommand{\arraystretch}{1.1}
\begin{adjustbox}{width=0.49\textwidth}
\begin{tabular}{lcccc}
\hline
\multirow{2}{*}{Model} & \multicolumn{2}{c}{Full context} & \multicolumn{2}{c}{Small context} \\
\cline{2-5}
 & Pass@1 & DIR & Pass@1 & DIR \\
\hline
phi-2 &	19.04 &	48.22 & 14.82 & 44.54 \\
phi-2$_{DepIT}$& \textbf{20.20} & \textbf{61.66} & \textbf{20.31} & \textbf{70.30}\\
\hline
StarCoder&	28.08 &	58.67 & 25.49 & 56.67 \\
StarCoder$_{DepIT}$&	\textbf{29.43}&	\textbf{69.80} & \textbf{28.73} & \textbf{71.48} \\
\hline
StarCoder2 &	27.77 &	60.57 & 23.27 & 53.49 \\
StarCoder2$_{DepIT}$& 	\textbf{28.45}&	\textbf{69.76} & \textbf{27.27} & \textbf{73.98} \\
\hline
CodeLlama &	38.65 &	62.26 & 35.66 & 62.67 \\
CodeLlama$_{DepIT}$& \textbf{38.85}&	\textbf{68.89} & \textbf{36.93} & \textbf{73.19} \\
\hline
\end{tabular}
\end{adjustbox}
\caption{Comparison of the performance of several models on \methodnamews after instruction tuning for dependency calls ($_{DepIT}$) with their pre-trained versions.}
\label{tab:dep_inst_tuning}
\end{table}
\section{Conclusion}

We propose an evaluation approach for repository-level code generation that rethinks the limitations of prior methods by assessing not only the functional correctness aspect but also dependency utilization to ensure code quality. We introduce \methodnamews, a novel Python code generation benchmark with executable capabilities, designed to evaluate the alignment of generated code with developer intent and correctness. Our experiments show that while pretrained LLMs excel in functional correctness, instruction-tuned models perform better in utilizing dependencies and debugging. However, existing models struggle to reuse provided dependencies, risking technical debt and code smells. We provide a comprehensive analysis of how LLMs leverage dependencies, revealing that context information significantly influences the results. Besides, we also introduce an instruction-tuning dataset that enhances dependency invocation accuracy and output correctness, even with limited context. Our contributions establish a foundation for future research in code generation, providing valuable evaluation techniques to drive the development of more capable and reliable models.


\section{Limitations}


In this work, we currently consider one level of dependency context, specifically the dependencies directly called from the target function. While this simplification facilitates manageable analysis and model development, it may not fully capture the valuable context necessary for leveraging models effectively. However, incorporating deep dependencies could significantly extend the input length, posing challenges in managing long context inputs and potentially exceeding the maximum input length. Our approach has demonstrated promising outcomes with the small-size context version, creating opportunities for integrating additional input context. Future research could explore incorporating multiple levels of dependencies, creating a more comprehensive graph that includes transitive dependencies, indirect calls, and broader contextual information. By doing so, we could enhance the model’s understanding of code interactions and improve its ability to handle intricate software execution scenarios.

\bibliography{custom}
\bibliographystyle{acl_natbib}

\newpage
\onecolumn
\appendix
\section*{{\huge Appendix}}

\section{Data Source}
\label{appedix_datasource}

Creating an executable benchmark within repository contexts poses significant challenges due to intricate setup requirements and the often inadequate installation instructions provided in repositories. Therefore, we select repositories that can be automatically set up using \texttt{pipreqs} to auto-detect the required packages. Additionally, to effectively utilize the test generation module of our pipeline, we select repositories that align with the literature on unit test generation. As a result, \methodnamews, is based on repositories from the unit test generation studies \citep{schafer2023adaptive, lukasczyk2023empirical, lemieux2023codamosa, liu2024no}, which provide executable environments and robust test generation capabilities. 

Some might claim that our dataset is constructed for based on existing works, thus, this approach can potentially lead to data leakage, especially when modern models are trained on similar datasets. If the benchmark relies heavily on known works, there’s a risk that the model may inadvertently learn from specific patterns or features present in those works, compromising its generalization ability. Despite this concern, experimental results demonstrate that even modern models struggle to handle the challenges posed by \methodnamews, indicating that the benchmark remains a valuable tool for assessing model performance. Besides, several models, including StarCoder2 and DeepSeek-Coder, have been pretrained using repository-level context. However, these models typically concatenate the contents of multiple files in a repository without filtering out irrelevant information or considering the selection of dependencies. This helps our dataset distinguish from the pretraining datasets of these models, thereby helping to mitigate the data leakage issue. To further measure the data leakage level compare to other benchmarks, we follow BenBench \cite{xu2024benchmarking} to report the perplexity and n-gram metrics, against three datasets—HumanEval, RepoCoder, and DevEval—analyzing their levels of data leakage when tested on three powerful LLMs. As indicated in Table \ref{tab:dataleak}, while HumanEval exhibits a significant level of data leakage, as expected, RepoExec demonstrates notably low levels of leakage. This is reflected in its higher perplexity and reduced n-gram overlap in most cases. The extent of data leakage in RepoExec is comparable to that of RepoCoder, a recently developed benchmark based on repositories created after January 1, 2022, to mitigate the data leakage problem. Consequently, the leakage issue in RepoExec is minimal, aligning with benchmarks like RepoCoder that utilize the latest repositories. We believe that repository-level code generation could be a valuable contribution to more robust and high-quality evaluation for advanced agent systems and dynamic environments \citep{huang2023agentcoder, phan2024hyperagent, yadav-etal-2023-exploring, hai2024continual, tran2024preserving}. By assessing models in a more comprehensive coding context, this approach enables a deeper evaluation of their ability to understand, integrate, and adapt code components effectively.

\begin{table*}
\centering
\renewcommand{\arraystretch}{1.1}
\begin{adjustbox}{width=\textwidth}
\begin{tabular}{l|cccc|cccc}
\hline
\multirow{2}{*}{Model} & \multicolumn{4}{c|}{perplexity} & \multicolumn{4}{c}{n-gram} \\
\cline{2-9}
& HumanEval&	RepoCoder&	DevEval&	RepoExec &	HumanEval&	RepoCoder&	DevEval&	RepoExec  \\
\hline
Mistral-7B-v0.3&	2.10&	\textbf{177.99}&	170.67&	120.77& 0.35&	0.30&	0.26&	\textbf{0.18}\\
\midrule
CodeLlama-7B-Python&	1.96&	\textbf{90.71}&	32.14&	72.77& 0.43&	\textbf{0.30}&	0.43&	\textbf{0.30} \\
\midrule
CodeLlama-13B-Python&	1.96&	72.66&	23.83&	\textbf{126.51}& 0.48&	\textbf{0.31}&	0.41&	\textbf{0.31} \\
\hline
\end{tabular}
\end{adjustbox}
\caption{Comparison of data leakage level between \methodnamews and other benchmarks.}
\label{tab:dataleak}
\end{table*}

\section{Evaluation Metrics: Match-Based vs. Execution-Based}

Several code generation benchmarks utilize match-based metrics like Edit similarity (ES), BLEU, and CodeBLEU for evaluation \citep{ding-etal-2024-cocomic-code, DBLP:conf/icml/ShrivastavaLT23, DBLP:conf/nips/DingWADTJRNBRX23, liao2023context, yin2018learning, iyer2018mapping}. These metrics are straightforward to apply and may exhibit a strong correlation with execution metrics such as Pass@k. However, they cannot accurately measure functional correctness. For instance, comparing two Python code snippets where the only difference is the ":" character could result in a good ES and BLEU score. Nevertheless, one snippet may contain a syntax error, highlighting a limitation in these metrics for assessing true functionality.

\begin{figure*}
  \makebox[\textwidth][c]{\includegraphics[width=\textwidth]{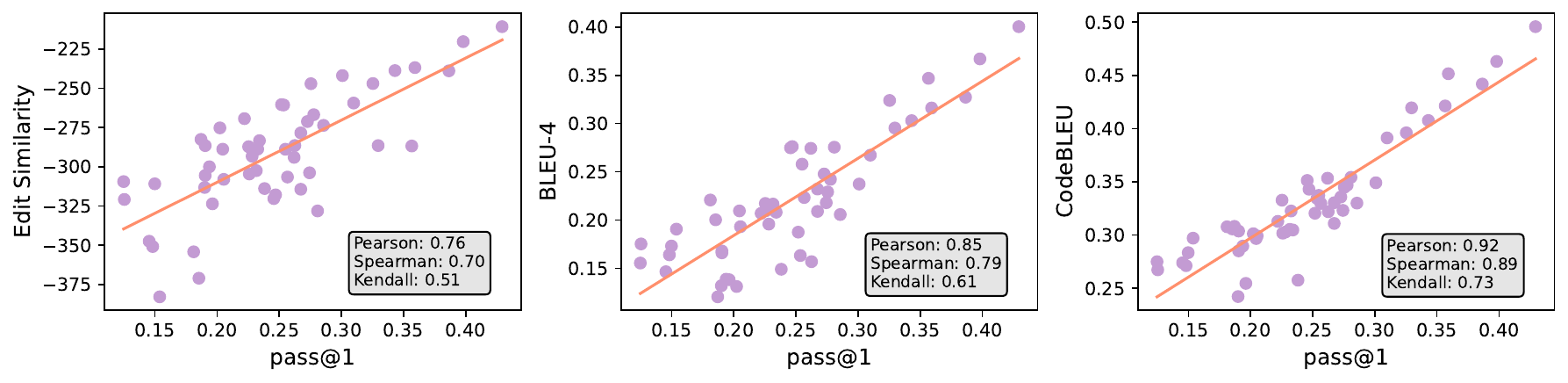}}
  \caption{Correlation between Match-based metrics and Execution-based metric (Pass@1).}
  \label{fig:correlation_between_metrics}
\end{figure*}

\begin{figure*}[h]
  \makebox[\textwidth][c]{\includegraphics[width=0.8\textwidth]{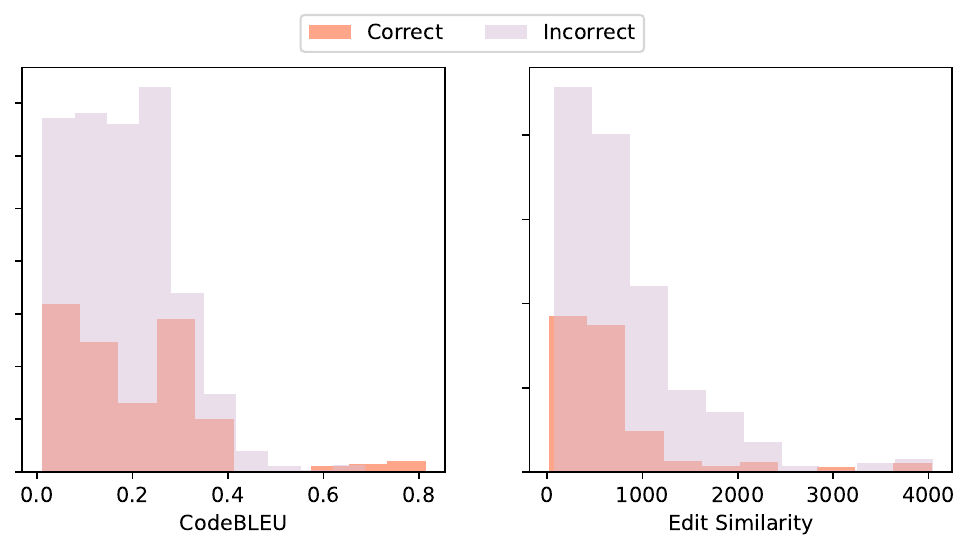}}
  \caption{Match-based metric distributions between Correct and Incorrect solutions}
  \label{fig:distribution_between_metrics}
\end{figure*}

Figure \ref{fig:correlation_between_metrics} demonstrates that these metrics on average can achieve a strong correlation with Pass@k, with CodeBLEU showing the highest correlation with Pearson score of 0.92. However, upon closer inspection of the score distribution between correct and incorrect solutions (Figure \ref{fig:distribution_between_metrics}), a considerable overlap becomes apparent. This underscores the limitation of match-based metrics in accurately measuring the correctness of code generation.

\section{Coverage Enhancement Effectiveness}
\label{appendix_coverage}

\begin{figure}[h]
  \makebox[\textwidth][c]{\includegraphics[width=0.95\textwidth]{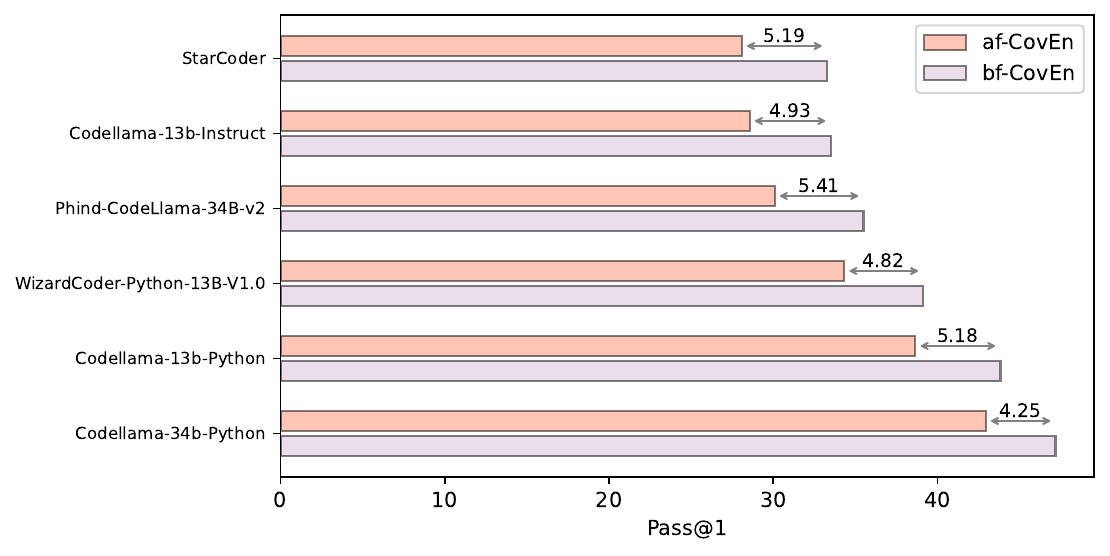}}
  \caption{Performance of various CodeLMs on \methodnamews before (bf-) and after (af-) Coverage Enhancement (CovEn) stage.}
  \label{fig:cov_performance}
\end{figure}

Weak unit tests may inadvertently allow incorrect implementations to be determined as correct. Even with human-written tests, the overlooked coverage rates lead to evaluations that are incomplete and potentially misleading \cite{liu2023is}. We present evidence supporting this argument, underscoring the limitations of prior work on code generation within repository-level contexts. As depicted in Figure \ref{fig:cov_performance}, enhancing the number of test cases and coverage rates leads to a significant increase in the identification of incorrect generated solutions, causing the Pass@1 score to drop markedly (by over 5\%). We investigated several solutions and found that most of the generated results did not fully utilize the given context (considered as human-provided). Instead, they primarily focused on addressing the problem described in the given natural language description. This indirectly overlooks the developer's intentions, such as testing edge or corner cases, highlighting the limitations in following and understanding the provided intent and dependency context in these models. In summary, these findings underscore the effectiveness and importance of maintaining high-quality test cases for evaluation purposes.

\section{Test case generation}
\label{appendix_testgen}

\paragraph{Generated test quality discussion:} To ensure generated tests align with the target function's requirements, we employ several quality control measures, as discussed in Section \ref{sec:testgen_tool}. First, we provide the full target function in the LLM prompt to guide the generation of tests with the intended input and output formats. We then apply a Syntax-based filter to confirm that the generated tests correctly call the function with its signature and parameters. Additionally, to verify the accuracy of the tests, we execute them to confirm their pass status and measure their line coverage. 

In some minor cases, weak tests or even wrong ones may still arise despite these measures. For instance, flaky tests could pass and achieve high coverage but produce inconsistent outputs due to randomness. To address this, we execute tests multiple times and exclude those with inconsistent results. Similarly, concerns about unintended function calls are mitigated by deduplicating test cases and ensuring diversity in how functions are invoked. This process yields an average of 99 test cases per problem (Table \ref{tab:dataset_attr}), ensuring robust evaluation even if some weaker tests are included. While we do not require such extensive coverage for correctness, this approach enhances the likelihood of strong test cases identifying implementation issues effectively. Furthermore, the test enhancement module is provided with a few examples of generated tests from the previous step to enhance coverage of edge cases and corner cases to ensure specific requirements.

\paragraph{Test geneneration prompts:} Below, we present two prompts designed for generating initial unit tests and improving the quality of existing tests.

\begin{tcolorbox}[colback=blue!5!white,colframe=gray!75!black,title=Initial Test Generation Prompt]

\lstset{basicstyle=\footnotesize\ttfamily}
\begin{lstlisting}
{function_under_test}
# test to check the correctness of "{function_name}" function
assert 
\end{lstlisting}
\end{tcolorbox}

\begin{tcolorbox}[colback=blue!5!white,colframe=gray!75!black,title=Coverage Enhancement Prompts]

\textbf{Prompt 1:} 

Here are some Python unit test functions and the focal function that they test: 

\# Test functions:

\{existing\_test\_functions\} 

\# Focal function:

\{function\_under\_test\}

Write more unit test functions that will increase the test coverage of the function under test.

-----------------------------------------------------------------------------------------------------

\textbf{Prompt 2:} 

Here are some Python unit test functions and the focal function that they test:

\# Test functions:

\{existing\_test\_functions\}

\# Focal function:

\{function\_under\_test\}

Write more unit test functions that will cover corner cases missed by the original and will increase the test coverage of the function under test.

-----------------------------------------------------------------------------------------------------

\textbf{Prompt 3:} 

Here is a focal function under test:

\{function\_under\_test\}

This function under test can be tested with these Python unit test functions: 

\{existing\_test\_functions\}

Here is an extended version of the unit test function that includes additional unit test cases that will cover methods, edge cases, corner cases, and other features of the function under test that were missed by the original unit test functions:
\end{tcolorbox}

\section{Data Formating}
\label{appendix_dataformat}
\subsection{BasePrompt}

\lstset{language=Python}
\begin{tcolorbox}[colback=blue!5!white,colframe=gray!75!black,title=Example of Full Context]

\begin{lstlisting}
import base64
import random
import unicodedata
import zlib
from typing import Union
from uuid import uuid4
from ._regex import *
from .errors import InvalidInputError
from .validation import is_snake_case, is_full_string, is_camel_case, is_integer, is_string

CAMEL_CASE_REPLACE_RE = re.compile(r'([a-z]|[A-Z]+)(?=[A-Z])')

class InvalidInputError(TypeError):
    """
    Custom error raised when received object is not a string as expected.
    """

    def __init__(self, input_data: Any):
        """
        :param input_data: Any received object
        """
        type_name = type(input_data).__name__
        msg = 'Expected "str", received "{}"'.format(type_name)
        super().__init__(msg)

def is_string(obj: Any) -> bool:
    """
    Checks if an object is a string.

    *Example:*

    >>> is_string('foo') # returns true
    >>> is_string(b'foo') # returns false

    :param obj: Object to test.
    :return: True if string, false otherwise.
    """
    return isinstance(obj, str)

def is_camel_case(input_string: Any) -> bool:
    """
    Checks if a string is formatted as camel case.

    A string is considered camel case when:

    - it's composed only by letters ([a-zA-Z]) and optionally numbers ([0-9])
    - it contains both lowercase and uppercase letters
    - it does not start with a number

    *Examples:*

    >>> is_camel_case('MyString') # returns true
    >>> is_camel_case('mystring') # returns false

    :param input_string: String to test.
    :type input_string: str
    :return: True for a camel case string, false otherwise.
    """
    return is_full_string(input_string) and CAMEL_CASE_TEST_RE.match(input_string) is not None
  
def camel_case_to_snake(input_string, separator='_'):
    """
    Convert a camel case string into a snake case one.
    (The original string is returned if is not a valid camel case string)

    *Example:*

    >>> camel_case_to_snake('ThisIsACamelStringTest') # returns 'this_is_a_camel_case_string_test'

    :param input_string: String to convert.
    :type input_string: str
    :param separator: Sign to use as separator.
    :type separator: str
    :return: Converted string.
    """
\end{lstlisting}
\end{tcolorbox}

\begin{tcolorbox}[colback=blue!5!white,colframe=gray!75!black,title=Example of Medium Context]

\begin{lstlisting}
import base64
import random
import unicodedata
import zlib
from typing import Union
from uuid import uuid4
from ._regex import *
from .errors import InvalidInputError
from .validation import is_snake_case, is_full_string, is_camel_case, is_integer, is_string

CAMEL_CASE_REPLACE_RE = re.compile(r'([a-z]|[A-Z]+)(?=[A-Z])')

class InvalidInputError(TypeError):
    """
    Custom error raised when received object is not a string as expected.
    """

    def __init__(self, input_data: Any):
        """
        :param input_data: Any received object
        """

def is_string(obj: Any) -> bool:
    """
    Checks if an object is a string.

    *Example:*

    >>> is_string('foo') # returns true
    >>> is_string(b'foo') # returns false

    :param obj: Object to test.
    :return: True if string, false otherwise.
    """

def is_camel_case(input_string: Any) -> bool:
    """
    Checks if a string is formatted as camel case.

    A string is considered camel case when:

    - it's composed only by letters ([a-zA-Z]) and optionally numbers ([0-9])
    - it contains both lowercase and uppercase letters
    - it does not start with a number

    *Examples:*

    >>> is_camel_case('MyString') # returns true
    >>> is_camel_case('mystring') # returns false

    :param input_string: String to test.
    :type input_string: str
    :return: True for a camel case string, false otherwise.
    """

def camel_case_to_snake(input_string, separator='_'):
    """
    Convert a camel case string into a snake case one.
    (The original string is returned if is not a valid camel case string)

    *Example:*

    >>> camel_case_to_snake('ThisIsACamelStringTest') # returns 'this_is_a_camel_case_string_test'

    :param input_string: String to convert.
    :type input_string: str
    :param separator: Sign to use as separator.
    :type separator: str
    :return: Converted string.
    """
\end{lstlisting}
\end{tcolorbox}

\begin{tcolorbox}[colback=blue!5!white,colframe=gray!75!black,title=Example of Small Context]

\begin{lstlisting}
import base64
import random
import unicodedata
import zlib
from typing import Union
from uuid import uuid4
from ._regex import *
from .errors import InvalidInputError
from .validation import is_snake_case, is_full_string, is_camel_case, is_integer, is_string

CAMEL_CASE_REPLACE_RE = re.compile(r'([a-z]|[A-Z]+)(?=[A-Z])')

class InvalidInputError(TypeError):

    def __init__(self, input_data: Any):

def is_string(obj: Any) -> bool:

def is_camel_case(input_string: Any) -> bool:

def camel_case_to_snake(input_string, separator='_'):
    """
    Convert a camel case string into a snake case one.
    (The original string is returned if is not a valid camel case string)

    *Example:*

    >>> camel_case_to_snake('ThisIsACamelStringTest') # returns 'this_is_a_camel_case_string_test'

    :param input_string: String to convert.
    :type input_string: str
    :param separator: Sign to use as separator.
    :type separator: str
    :return: Converted string.
    """
\end{lstlisting}
\end{tcolorbox}

\subsection{InstructPrompt}

\begin{tcolorbox}
[colback=blue!5!white,colframe=gray!75!black,title=Instruction Prompt Templates]

\lstset{basicstyle=\footnotesize\ttfamily, language={}}

\textbf{Prompt 1:}
\begin{lstlisting}
### Instruction:

Write a Python function `{target_function_signature}` to solve the following problem:
{target_function_docstring}

### Response:
{BasePrompt}
\end{lstlisting}
-----------------------------------------------------------------------------------------------------
\textbf{Prompt 2:}
\begin{lstlisting}
### Instruction:

{dependency_context}
The provided code snippet includes necessary dependencies for implementing the `{target_function_name}` function. Write a Python function `{target_function_signature}` to solve the following problem:
{target_function_docstring}

### Response:
{target_function_prompt}

\end{lstlisting}
\end{tcolorbox}

\lstset{basicstyle=\scriptsize\ttfamily, language={}}
\begin{tcolorbox}[colback=blue!5!white,colframe=gray!75!black,title=Example of Prompt 1 for Small Context]
\begin{lstlisting}
### Instruction:

Write a Python function `camel_case_to_snake(input_string, separator='_')` to solve the following problem:
"""
Convert a camel case string into a snake case one.
(The original string is returned if is not a valid camel case string)

*Example:*

>>> camel_case_to_snake('ThisIsACamelStringTest') # returns 'this_is_a_camel_case_string_test'

:param input_string: String to convert.
:type input_string: str
:param separator: Sign to use as separator.
:type separator: str
:return: Converted string.
"""

### Response:
import base64
import random
import unicodedata
import zlib
from typing import Union
from uuid import uuid4
from ._regex import *
from .errors import InvalidInputError
from .validation import is_snake_case, is_full_string, is_camel_case, is_integer, is_string

CAMEL_CASE_REPLACE_RE = re.compile(r'([a-z]|[A-Z]+)(?=[A-Z])')

class InvalidInputError(TypeError):

    def __init__(self, input_data: Any):

def is_string(obj: Any) -> bool:

def is_camel_case(input_string: Any) -> bool:

def camel_case_to_snake(input_string, separator='_'):
    """
    Convert a camel case string into a snake case one.
    (The original string is returned if is not a valid camel case string)

    *Example:*

    >>> camel_case_to_snake('ThisIsACamelStringTest') # returns 'this_is_a_camel_case_string_test'

    :param input_string: String to convert.
    :type input_string: str
    :param separator: Sign to use as separator.
    :type separator: str
    :return: Converted string.
    """
\end{lstlisting}
\end{tcolorbox}

\begin{tcolorbox}[colback=blue!5!white,colframe=gray!75!black,title=Example of Prompt 2 for Small Context]

\begin{lstlisting}
### Instruction

import base64
import random
import unicodedata
import zlib
from typing import Union
from uuid import uuid4
from ._regex import *
from .errors import InvalidInputError
from .validation import is_snake_case, is_full_string, is_camel_case, is_integer, is_string

CAMEL_CASE_REPLACE_RE = re.compile(r'([a-z]|[A-Z]+)(?=[A-Z])')

class InvalidInputError(TypeError):

    def __init__(self, input_data: Any):

def is_string(obj: Any) -> bool:

def is_camel_case(input_string: Any) -> bool:

The provided code snippet includes necessary dependencies for implementing the `camel_case_to_snake` function. Write a Python function `camel_case_to_snake(input_string, separator='_')` to solve the following problem:
"""
Convert a camel case string into a snake case one.
(The original string is returned if is not a valid camel case string)

*Example:*

>>> camel_case_to_snake('ThisIsACamelStringTest') # returns 'this_is_a_camel_case_string_test'

:param input_string: String to convert.
:type input_string: str
:param separator: Sign to use as separator.
:type separator: str
:return: Converted string.
"""

### Response:
def camel_case_to_snake(input_string, separator='_'):
    """
    Convert a camel case string into a snake case one.
    (The original string is returned if is not a valid camel case string)

    *Example:*

    >>> camel_case_to_snake('ThisIsACamelStringTest') # returns 'this_is_a_camel_case_string_test'

    :param input_string: String to convert.
    :type input_string: str
    :param separator: Sign to use as separator.
    :type separator: str
    :return: Converted string.
    """
\end{lstlisting}
\end{tcolorbox}

\section{Studied LLMs: Supplemental results}
\label{appendix_llms_exp}
In this section, we offer supplementary results from the evaluation of LLMs on \methodnamews. Table \ref{tab:codegen} presents the Dependency Invocation Rate (DIR) for the experimented LLMs. When comparing models of the same size, it is shown that instruction-tuned models more effectively follow human intent in utilizing the provided dependencies with InstructPrompt. For example, WizardCoder outperforms CodeLlama by 5\%, and the instruction-tuned version of Mixtral-8x7B shows a 10\% improvement over its foundation version. This highlights the strong capability of instruction-tuned models to follow the given context effectively. Besides, using the Medium context leads to a significant decline in both Pass@k and DIR using BasePrompt. This implies the generation of empty function bodies using this template. 

\begin{figure}[t]
  \makebox[\textwidth][c]{\includegraphics[width=0.85\textwidth]{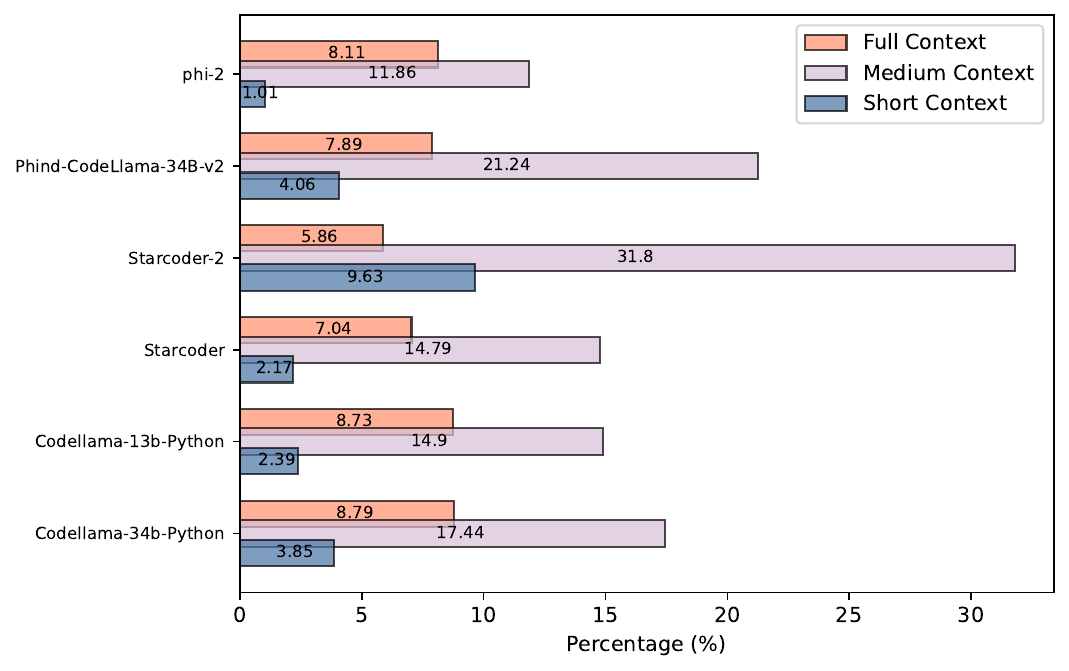}}
  \caption{Percentage of generated outputs that result in empty functions across various context types.}
  \label{fig:empty_generated}
\end{figure}

Indeed, Figure \ref{fig:empty_generated} illustrates the proportion of generated functions that are empty for each LLM using BasePrompt. The findings indicate that utilizing Medium context results in a substantial number of empty functions. This may be due to the input format of the context when using BasePrompt, which can mislead the model into interpreting dependency functions as few-shot examples. In the Medium context, the function bodies of dependencies are removed, making their format identical to the target function prompt. This similarity can mislead the LMs, resulting in empty solutions. Particularly, Starcoder-2 is heavily impacted by this issue, as over 31\% of its generated results are empty functions, revealing a significant weakness of the model. Meanwhile, small context effectively decreases the occurrence of empty function generation by the model and, in certain instances, improves models' ability in dependency calls (e.g. CodeLlama-13b-Python, WizardCoder-Python-13B-V1.0, and Mixtral-8x7B-Instruct-v0.1 in Table \ref{tab:codegen}). We believe that the following reasons could contribute to this observation. Firstly, employing small context reduces the input token count, preventing truncation when exceeding the maximum length limit, thus allowing uninterrupted solution generation by the model. This reduced context enables models to concentrate exclusively on dependency signatures, thereby enhancing the probability that generated solutions effectively utilize these dependency token names. Moreover, function names hold substantial semantic value by delineating the function's purpose. Many studies have underscored that code summarization heavily relies on extracting information from function names \citep{haldar2024analyzing, mondal-etal-2023-robust, sontakke2022code}. Therefore, this concise representation of dependencies has the potential to improve how models utilize dependencies in generating code.

Additionally, Table \ref{tab:examples_gen} presents examples that support our findings in Section \ref{sec:llm_exp}. For instance, in the first example, we observe that the instruction-tuned model can effectively utilize the given dependencies to manage edge cases to raise an error, whereas the pretrained model fails to do so. This supports our first findings. Meanwhile, the second and third examples demonstrate that pretrained models try to reimplement or devise workarounds instead of leveraging the available context. Additionally, in the second example, the instruction-tuned model successfully identifies the relevant dependencies; however, it complicates the situation and fails to produce the correct solution. This may suggest that hallucinations complicate the outputs generated by instruction-tuned models. In the third example, both types of models successfully pass the tests. However, although the context provides the dependency for creating a new \texttt{Future} object, these models attempt to reimplement the \texttt{\_create\_future} function but fail to optimize memory usage effectively. This observation implies the potential of code smell and technical debt in these generated codes.

Besides, we present supplementary results for both closed-source and open-source models in Table \ref{tab:leaderboard}. Specifically, we employ greedy decoding and report pass@1 and DIR metrics, selecting the best-performing results for each model based on either BasePrompt or InstructPrompt with Full context. The results demonstrate strong performance in both pass@k and DIR metrics with advanced reasoning models like DeepSeek-R1 and GPT-4o, highlighting their reliability in code generation. However, all pass@1 results remain below 50\%, underscoring the challenges posed by \methodnamews.

\begin{table*}[t]
\footnotesize
\centering
\renewcommand{\arraystretch}{1.1}
\begin{adjustbox}{width=0.8\textwidth}
\begin{tabular}{l|l|l|c|c}
\toprule
ID & Model & Size (B) & Pass@1 & DIR \\
\midrule
1 & DeepSeek-R1 \cite{guo2025deepseek} & 671 & \textbf{42.57}& 70.86   \\
2 & DeepSeek-V3 \cite{liu2024deepseek} & 671 & \underline{42.00} & \textbf{80.35}   \\
3 & Llama 3.1-Instruct \cite{dubey2024llama} &  405 & 39.43 &	\underline{73.26} \\
4 & GPT-4o    & - & 38.00 & 	68.15   \\
5 & QwQ-Preview \cite{qwq-32b-preview} & 32 &  37.43&  57.07 \\
6 & DeepSeek-Coder-Instruct	\cite{guo2024deepseek} & 33 & 35.71&	67.02 \\
7 & Yi-1.5 \cite{young2024yi} & 34&	35.43&	61.34 \\
8 & Qwen-2.5-Coder-Instruct \cite{hui2024qwen2} & 14 & 34.28 & 65.68 \\
9 & CodeQwen1.5 \cite{qwen} &7	& 32.29&	59.48 \\
10 & OpenCodeInterpreter \cite{zheng2024opencodeinterpreter}& 33 &	31.42&	61.28 \\
11 & Phi-4 \cite{abdin2024phi} & 14 & 30.86& 71.10 \\
12 & GPT-4o-mini& - & 30.29& 74.75   \\
13 & Llama 3.3-Instruct \cite{dubey2024llama}&	70 & 29.43 & 67.20 \\
14 & DeepSeek-Coder \cite{guo2024deepseek} &6.7 & 	28.86&	57.44 \\
15 & Qwen-2.5-Instruct \cite{yang2024qwen2} & 14 &	28.85&	65.63\\
16 & starcoder2 \cite{lozhkov2024starcoder} &15 &	28.57&	60.97 \\
17 & Gemma2 \cite{team2024gemma} & 27 & 28.00 & 71.68 \\
18 & Llama-3.1-Instruct \cite{dubey2024llama}& 70 & 25.14 & 66.95 \\

\bottomrule
\end{tabular}
\end{adjustbox}
\caption{Performance of various LLMs on \methodnamews.}
\label{tab:leaderboard}
\end{table*}

\section{Multi-round Debugging}
\label{appendix_debug}
\begin{tcolorbox}[colback=blue!5!white,colframe=gray!75!black,title=Prompt for Debugging]
\lstset{basicstyle=\footnotesize\ttfamily, language={}}
\begin{lstlisting}
{dependency_context}
# The provided code snippet includes necessary dependencies for implementing the `{target_function_name}` function. Write a Python function `{target_function_signature}` to solve the following problem:
{target_function_docstring}

# Here is the current solution.
{error_solution}

# When executing the below test case.
{failed_test_case}

# The provided python code solution fails the test with the following errors, please correct them.
{error_log}

# Please provide the modified code for me to review and provide feedback.
{target_function_prompt}

\end{lstlisting}
\end{tcolorbox}

We employ Multi-round debugging in code generation, which iteratively refines and improves the generated code through multiple cycles of debugging. Following the execution of unit tests on the generated functions, we extract the error log if the code fails to run. We employ the following prompt to utilize the model for bug fixing. This process is iterated multiple times until either the correct code is achieved or the maximum number of rounds is reached. Specifically, we set the maximum number of rounds to 3 and experimented on three models WizardCoder, GPT-3.5 and CodeLlama-13b-Python.

\begin{figure}[h]
  \makebox[\textwidth][c]{\includegraphics[width=0.85\textwidth]{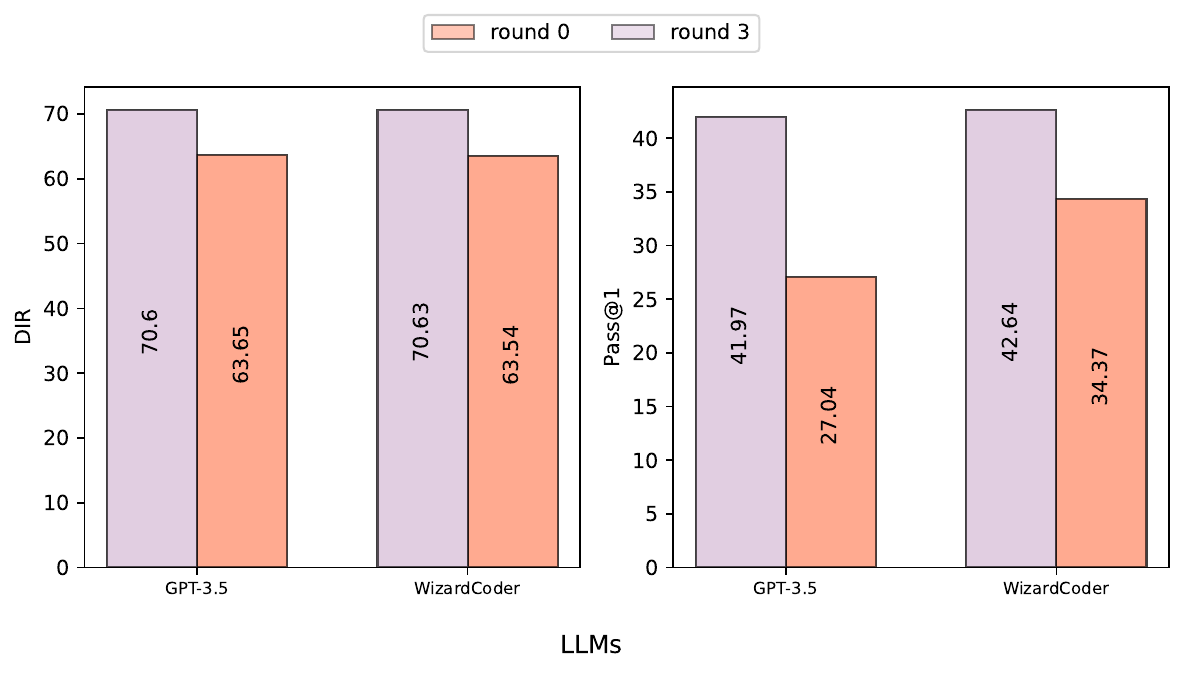}}
  \caption{Improvement of instruction-tuning models on Pass@1 and DIR after 3-round debugging process.}
  \label{fig:instruction_debugrs}
\end{figure}

Table \ref{tab:debug} shows the improvement across three rounds of debugging in various models. We observe that GPT-3.5 and WizardCoder demonstrate a high capacity for debugging with improvement of over 10\% and 7\% in Pass@1, respectively, while CodeLlama fails to take advantage of this process. Additionally, the DIR has also shown a significant improvement (over 7\%) after three rounds of debugging in these two instruction models (Figure \ref{fig:instruction_debugrs}). These findings indicate a promising approach using self-refinement with debugging for code generation, which can enhance both the correctness and the utilization of given dependencies.

\begin{figure}[t]
  \makebox[\textwidth][c]{\includegraphics[width=
\textwidth]{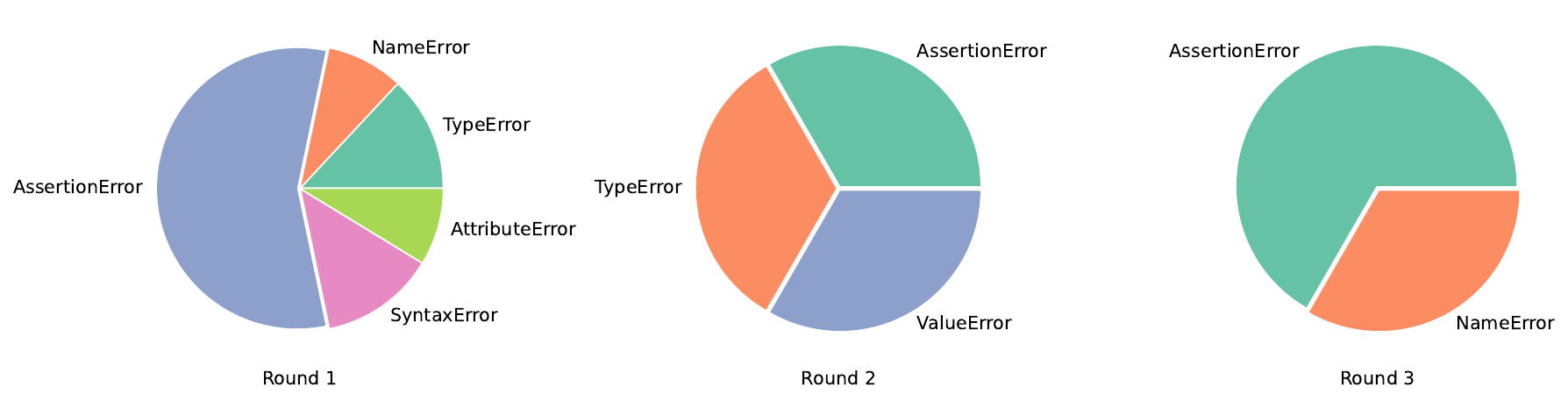}}
  \caption{Fixed error types of WizardCoder across 3 rounds of the debugging process.}
  \label{fig:fixed_error}
\end{figure}

We also present data on the number of error types corrected in each round of WizardCoder, as illustrated in Figure \ref{fig:fixed_error}. We can see that \texttt{AssertionError} makes up the majority of errors across all rounds. This error type indicates either incorrect outputs from the generated code or the presence of empty function bodies (return None). However, by incorporating the test output guide, the model effectively addressed most of these errors. Furthermore, fundamental issues like \texttt{SyntaxError} or \texttt{AttributeError} were promptly rectified during the initial round.

\section{Context length, model size and families analysis}

\begin{figure}[t]
  \makebox[\textwidth][c]{\includegraphics[width=\textwidth]{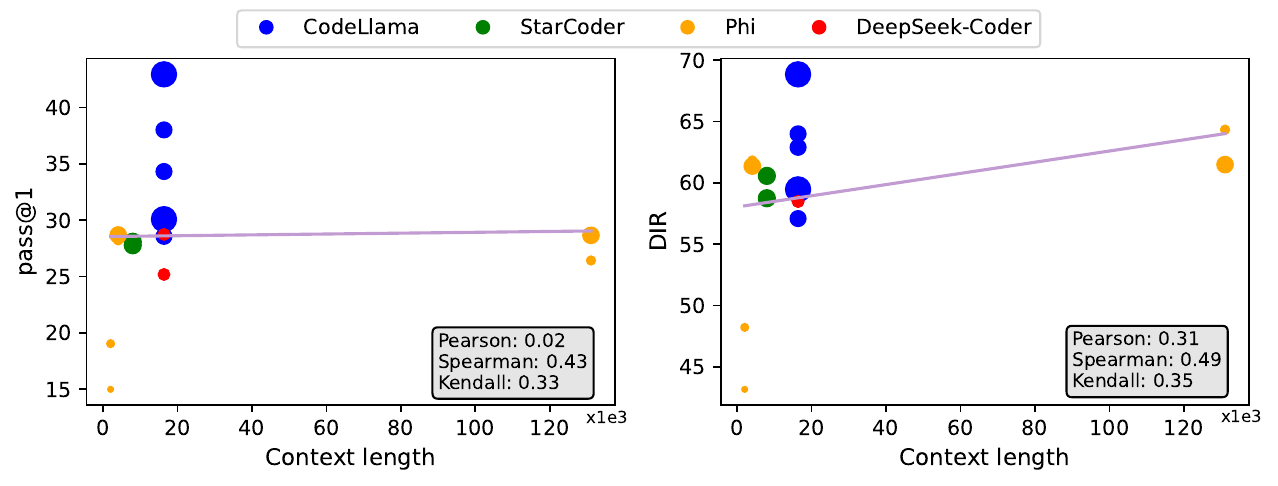}}
  \caption{Correlation of context length to the model performance on \methodnamews. The size of the dots indicates the model size, while the color represents the model family.}
  \label{fig:context_lenght}
\end{figure}

Long-context models can enhance the ability to comprehend and select relevant context from lengthy inputs to effectively solve the required task.  In this section, we examine how the context length supported by each model affects their performance in \methodnamews. Figure \ref{fig:context_lenght} demonstrates the relationship between support context length, model size, and model family in relation to pass@k and DIR scores.

We observe that context length has a weak correlation with model performance on our dataset. In contrast, model size (scaling law) and model family, which encompass different training methods (pretraining or instruction tuning), training datasets, and architectures, show a more significant impact. The weak correlation with context length can be explained by our approach's ability to already capture relevant information (e.g. dependency) for each data sample, resulting in the pruning of context length for practical usage (363 tokens on average shown in Table \ref{tab:dataset_attr}). Meanwhile, models that support long contexts are often trained on data containing a mix of relevant and irrelevant information and are evaluated on their ability to retrieve the correct context in a needle-in-a-haystack scenario \citep{roziere2023code, ivgi2023efficient, liu2024lost}. Therefore, models with varying context lengths might show a weak correlation to performance in our scenario.

\section{Dependency Extraction Tool}
\label{sec:appendix_tool}

We present \texttt{pydepcall}, a Python library designed to extract function dependencies from any repository.  Our tool provides a quick, practical solution for scalability, requiring only three lines of code to extract dependencies for all functions in any repository. Specifically, our tool is based on two conventions for retrieving function dependencies of a programming language:
\begin{itemize}
    \item \textbf{Import Convention (for cross-file dependencies):} The convention mentions how a programming language imports modules within a repository. This can be utilized to extract dependency names from different files. For instance, in Python, this can be done using statements such as "\texttt{from file import function/class/variable}" or "\texttt{from file1.file2 import function/class/variable.}"
    \item \textbf{Calling Dependencies Convention:} This principle pertains to how a program invokes variables, functions, and classes. In Python, functions are typically called by specifying their names followed by parentheses (e.g., \texttt{function1(param1, param2))}, or by using a dot notation to access attributes or methods from a module (e.g., \texttt{file1\textbf{.}function1, class1\textbf{.}attribute1}).
\end{itemize}

Besides, \texttt{pydepcall} can extract dependencies at various depths, up to 100 levels deep, which can help the community explore the integration of a deeper context. We provide a brief overview of its usage in the following code snippet.

\lstset{language=Python}
\begin{tcolorbox}[colback=blue!5!white,colframe=gray!75!black,title=\texttt{pydepcall} usage]

\begin{lstlisting}
from pydepcall import Extractor

# If you want to extract all module files in the repository
reposrc = YOUR_LOCAL_PATH_OF_REPO
extractor = Extractor(reposrc)
output = extractor.extract()

# If you want to extract a specific module file in the repository
reposrc = YOUR_LOCAL_PATH_OF_REPO
module_file = YOUR_LOCAL_PATH_OF_FILE_IN_REPO
extractor = Extractor(reposrc, module_file)
output = extractor.extract()
\end{lstlisting}
\end{tcolorbox}

\newpage
\begingroup

\lstset{language=Python}

\centering
\begin{longtable}{|p{0.45\textwidth}|p{0.45\textwidth}|}
\hline
\textbf{Input Prompt \& Solution} & \textbf{Generation output}\\ 
\hline \hline
\begin{lstlisting}
# Prompt: 
import base64
import random
import unicodedata
...

class InvalidInputError(TypeError):
    ...
    
def is_string(obj: Any) -> bool:
    ...

def reverse(input_string: str) -> str:
    """
    Returns the string with its chars reversed.

    *Example:*

    >>> reverse('hello') # returns 'olleh'

    :param input_string: String to revert.
    :type input_string: str
    :return: Reversed string.
    """
\end{lstlisting}

\begin{lstlisting}
--------------------------------------------------
# Solution:
def reverse(input_string: str) -> str:
    """
    Returns the string with its chars reversed.

    *Example:*

    >>> reverse('hello') # returns 'olleh'

    :param input_string: String to revert.
    :type input_string: str
    :return: Reversed string.
    """
    if not is_string(input_string):
        raise InvalidInputError(input_string)

    return input_string[::-1]
\end{lstlisting}

& 
\begin{lstlisting}
# Model type: Pretrained
# Passed: No
def reverse(input_string: str) -> str:
    """
    Returns the string with its chars reversed.

    *Example:*

    >>> reverse('hello') # returns 'olleh'

    :param input_string: String to revert.
    :type input_string: str
    :return: Reversed string.
    """
    return input_string[::-1]
\end{lstlisting}

\begin{lstlisting}
--------------------------------------------------
# Model type: Instruction-tuned
# Passed: Yes
def reverse(input_string: str) -> str:
    """
    Returns the string with its chars reversed.

    *Example:*

    >>> reverse('hello') # returns 'olleh'

    :param input_string: String to revert.
    :type input_string: str
    :return: Reversed string.
    """
    if not is_string(input_string):
        raise InvalidInputError(input_string)
    return input_string[::-1]
\end{lstlisting}
\\ \hline \hline
\begin{lstlisting}
# Prompt:
import random
import luigi
import luigi.configuration
...

class hadoopcli(luigi.Config):
    command = ...
    version = ...
    
def get_configured_hadoop_version():
    """
    CDH4 (hadoop 2+) has a slightly different syntax for interacting with hdfs
    via the command line.

    The default version is CDH4, but one can override
    this setting with "cdh3" or "apache1" in the hadoop section of the config
    in order to use the old syntax.
    """
--------------------------------------------------
# Solution:
def get_configured_hadoop_version():
    """
    CDH4 (hadoop 2+) has a slightly different syntax for interacting with hdfs
    via the command line.

    The default version is CDH4, but one can override
    this setting with "cdh3" or "apache1" in the hadoop section of the config
    in order to use the old syntax.
    """
    return hadoopcli().version.lower()
\end{lstlisting}
& 

\begin{lstlisting}
# Model type: Pretrained
# Passed: Yes
def get_configured_hdfs_client():
    """
    This is a helper that fetches the configuration value for 'client' in
    the [hdfs] section. It will return the client that retains backwards
    compatibility when 'client' isn't configured.
    """
    return luigi.configuration.get_config().get('hdfs', 'client', 'hadoopcli')

--------------------------------------------------
# Model type: Instruction-tuned
# Passed: No
def get_configured_hadoop_version():
    """
    CDH4 (hadoop 2+) has a slightly different syntax for interacting with hdfs
    via the command line.

    The default version is CDH4, but one can override
    this setting with "cdh3" or "apache1" in the hadoop section of the config
    in order to use the old syntax.
    """
    hadoop_cli = hadoopcli().command.split()
    hadoop_version = hadoopcli().version
    if hadoop_version == "cdh3":
        return hadoop_cli + ["fs"]
    elif hadoop_version == "apache1":
        return hadoop_cli + ["dfs"]
    else:
        return hadoop_cli

\end{lstlisting}
\\ \hline\hline
\begin{lstlisting}
# Prompt:
import asyncio
import builtins
import collections
...

def is_future(x: Any) -> bool:
    return isinstance(x, FUTURES)


def _create_future() -> Future:
    future = Future()  # type: Future
    # Fixup asyncio debug info by removing extraneous stack entries
    source_traceback = getattr(future, "_source_traceback", ())
    while source_traceback:
        # Each traceback entry is equivalent to a
        # (filename, self.lineno, self.name, self.line) tuple
        filename = source_traceback[-1][0]
        if filename == __file__:
            del source_traceback[-1]
        else:
            break
    return future

def maybe_future(x: Any) -> Future:
    """Converts ``x`` into a `.Future`.

    If ``x`` is already a `.Future`, it is simply returned; otherwise
    it is wrapped in a new `.Future`.  This is suitable for use as
    ``result = yield gen.maybe_future(f())`` when you don't know whether
    ``f()`` returns a `.Future` or not.

    .. deprecated:: 4.3
    This function only handles ``Futures``, not other yieldable objects.
    Instead of `maybe_future`, check for the non-future result types
    you expect (often just ``None``), and ``yield`` anything unknown.
    """
--------------------------------------------------
# Solution:
def maybe_future(x: Any) -> Future:
    """Converts ``x`` into a `.Future`.

    If ``x`` is already a `.Future`, it is simply returned; otherwise
    it is wrapped in a new `.Future`.  This is suitable for use as
    ``result = yield gen.maybe_future(f())`` when you don't know whether
    ``f()`` returns a `.Future` or not.

    .. deprecated:: 4.3
       This function only handles ``Futures``, not other yieldable objects.
       Instead of `maybe_future`, check for the non-future result types
       you expect (often just ``None``), and ``yield`` anything unknown.
    """
    if is_future(x):
        return x
    else:
        fut = _create_future()
        fut.set_result(x)
        return fut
\end{lstlisting}
& 

\begin{lstlisting}
# Model type: Pretrained
# Passed: Yes
def maybe_future(x: Any) -> Future:
    """Converts ``x`` into a `.Future`.

    If ``x`` is already a `.Future`, it is simply returned; otherwise
    it is wrapped in a new `.Future`.  This is suitable for use as
    ``result = yield gen.maybe_future(f())`` when you don't know whether
    ``f()`` returns a `.Future` or not.

    .. deprecated:: 4.3
    This function only handles ``Futures``, not other yieldable objects.
    Instead of `maybe_future`, check for the non-future result types
    you expect (often just ``None``), and ``yield`` anything unknown.
    """
    if is_future(x):
        return x
    future = Future()  # type: Future
    future.set_result(x)
    return future

--------------------------------------------------
# Model type: Instruction-tuned
# Passed: Yes
def maybe_future(x: Any) -> Future:
    """Converts ``x`` into a `.Future`.

    If ``x`` is already a `.Future`, it is simply returned; otherwise
    it is wrapped in a new `.Future`.  This is suitable for use as
    ``result = yield gen.maybe_future(f())`` when you don't know whether
    ``f()`` returns a `.Future` or not.

    .. deprecated:: 4.3
    This function only handles ``Futures``, not other yieldable objects.
    Instead of `maybe_future`, check for the non-future result types
    you expect (often just ``None``), and ``yield`` anything unknown.
    """
    if isinstance(x, Future):
        return x
    elif isawaitable(x):
        return asyncio.ensure_future(x)
    else:
        future = Future()
        future.set_result(x)
        return future

\end{lstlisting}
\\
\hline
\caption{Output samples of Pretrained and Instruction-tuned models.}
\label{tab:examples_gen}
\end{longtable}

\endgroup

\end{document}